\documentclass[aps,prx,reprint,superscriptaddress,floatfix,amsmath,amssymb,longbibliography]{revtex4-2}
\usepackage{epsfig}
\usepackage{graphics}
\usepackage{float} 
\usepackage{array}
\usepackage{multirow}
\usepackage{combelow}
\usepackage{yfonts}
\usepackage[normalem]{ulem}
\usepackage{graphicx}
\usepackage{amssymb}
\usepackage{mathrsfs}
\usepackage{bbm}
\usepackage{epsfig}
\usepackage{makecell}
\usepackage[left]{lineno}

\usepackage{bbm}
\usepackage{color}

\newcommand{\la}[1]{\label{#1}}

\newcommand{\be}{\begin{eqnarray}}
\newcommand{\ee}{\end{eqnarray}}

\usepackage{xcolor}
\usepackage{ulem}
\definecolor{purple}{rgb}{0.8,0,0.6}

\allowdisplaybreaks

\usepackage{hyperref}

\begin{document}
\begin{flushleft}
\end{flushleft}

\title{
\Large 
Non-Abelian Geometric Phases in Triangular Structures \\ And  Universal SU(2) Control in Shape Space
}

\author{J. Dai}
\email{jindai@bit.edu.cn}
\affiliation{Center for Quantum Technology Research, Key Laboratory of Advanced Optoelectronic Quantum 
Architecture and Measurements (MOE), School of Physics, Beijing Institute of Technology, Beijing 100081, China}
\author{A. Molochkov}
\email{molochkov.alexander@gmail.com}
\affiliation{Beijing Institute of Mathematical Sciences and Applications, Tsinghua University, 101408, Huairou District, Beijing, China}
\author{A.J. Niemi}
\email{Antti.Niemi@su.se}
\affiliation{Nordita, Stockholm University, Roslagstullsbacken 23, SE-106 91 Stockholm, Sweden}
\affiliation{Wilczek Quantum Center,  Shanghai Institute for Advanced Studies, University of Science and Technology of China,
 Shanghai 201315, China}
\author{J. Westerholm}
\email{Jan.Westerholm@abo.fi}
\affiliation{Faculty of Science and Engineering, \AA bo Akademi University, Vattenborgsv\"agen 3, FI-20500 \AA bo, Finland}

\begin{abstract}
We construct  holonomic quantum gates for qubits that are encoded in the near-degenerate vibrational $E$-doublet of a 
deformable  three-body system. Using Kendall's shape theory, we derive the Wilczek--Zee connection  that
governs adiabatic transport within the $E$-manifold. We show that its restricted holonomy group is $\mathrm{SU}(2)$, 
implying universal single-qubit control by closed loops in shape space. We provide explicit loops implementing 
a $\pi/2$ phase gate and a Hadamard-type gate. For two-qubit operations, we outline how linked holonomic 
cycles in arrays generate a controlled Chern--Simons phase, enabling an entangling controlled-$X$ (CNOT) 
gate.  We present a Ramsey/echo interferometric protocol that measures the Wilson loop trace of the Wilczek--Zee connection
for a control cycle, providing a gauge-invariant signature of the non-Abelian holonomy.
As a physically realizable demonstrator, we propose bond-length modulations of a Cs($6s$)--Cs($6s$)--Cs($nd_{3/2}$)
 Rydberg trimer in optical tweezers and specify operating conditions that suppress leakage out of the $E$-manifold.
\end{abstract}

\maketitle

\section{Introduction}  
Non-Abelian geometric phases~\cite{WilczekZee1984} provide one of the
conceptual foundations of holonomic quantum computation. In its original
form, the approach was formulated in terms of adiabatic loops of degenerate
eigenspaces~\cite{Zanardi1999}. The ensuing gate is holonomic: it is determined
by the closed path traced by the eigenspace and by the associated
Wilczek--Zee connection. It is therefore independent of the detailed timing
with which the path is traversed, provided the evolution remains adiabatic.
This geometric character can provide robustness against certain classes of
control imperfections. The framework has since been extended beyond the
original adiabatic setting to nonadiabatic schemes~\cite{Xu2012}, including
experimental implementations~\cite{Feng2013}. In recent years, geometric-gate
control has developed rapidly across several quantum-computing platforms. For
example, shortened-path nonadiabatic schemes have been proposed and
experimentally implemented in superconducting circuits~\cite{Li2021,Yang2023}.
Noncyclic geometric gates have been shown in trapped-ion settings to combine
speed with noise resilience~\cite{Zhang2021}. Semiconductor-spin experiments
have demonstrated high-fidelity geometric manipulation in germanium quantum
dots~\cite{Zhou2025}. Together, these advances show that geometric phases have
become a practical control resource for quantum information
processing~\cite{Zhang2023}.

As quantum information science advances across multiple hardware
platforms~\cite{Monroe1995,Gasparoni2004,ClarkeWilhelm2008,Devoret2013}, there
is growing interest in implementing geometric gates in scalable architectures
with long coherence times. Programmable arrays of neutral Rydberg atoms are
particularly promising. They combine uniform atomic qubits, high-fidelity
site-resolved laser control, and engineered
interactions~\cite{SaffmanRMP2010,BrowaeysLahaye2020}. In this setting,
ultralong-range Rydberg excitations can bind one Rydberg atom to two
ground-state atoms, forming triatomic molecular
states~\cite{Bendkowsky2010,Greene2000}. These Rydberg trimers realize
genuinely three-body, non-additive interactions, which have been established
spectroscopically~\cite{Fey2019}. Trimer spectral lines have also been used to
probe nonlocal three-body correlations and to identify new, tunable trimer
families~\cite{Kanungo2023,Bosworth2023,Hollerith2024}. These results indicate
that molecular trimers can serve as experimentally accessible units for
assembly into programmable arrays. This motivates our investigation of qubit
encodings in vibrational geometry and holonomic control through non-Abelian
geometric phases~\cite{WilczekZee1984,Xu2012,Feng2013}.

Complementing existing holonomic-gate proposals, we develop a framework in
which a qubit is encoded directly in the shape degrees of freedom of a physical
triangular system, such as a molecular trimer. We identify the logical
two-level system with a near-degenerate vibrational \(E\)-doublet of the
trimer. Kendall's shape sphere~\cite{Kendall-1999} then provides the geometric
space of triangular shapes on which this doublet is transported. Our central
result is the construction of the corresponding Wilczek--Zee
connection~\cite{WilczekZee1984,Zanardi1999}, which exposes the
\(\mathrm{SU}(2)\) gauge structure generated by vibrational shape dynamics of
triangular systems. Although our construction builds on adiabatic
Wilczek--Zee holonomy, its geometric structure also connects naturally to
nonadiabatic holonomic-gate schemes, where finite-time cyclic evolution of a
computational subspace is engineered using pulse-controlled few-level
systems~\cite{Sjoqvist2012,Xu2012,Feng2013,Abdumalikov2013}.

With the connection in hand, we show that the restricted holonomy group is
\(\mathrm{SU}(2)\). Hence, suitable closed loops in shape space can generate
universal single-qubit holonomic operations on the \(E\)-doublet. We introduce
the gauge-invariant trace of the Wilson loop as a compact and experimentally
accessible diagnostic of the non-Abelian shape-space holonomy. Finally, we
outline how linked deformation cycles in arrays of triangular units can
generate controlled geometric phases with a Chern--Simons description. When
the acquired phase depends on the joint logical state of two units, such linked
cycles provide a possible route to two-qubit entangling operations.

Triangular Rydberg--perturber configurations may realize the required
\(E\)-doublet through anisotropic electron--atom scattering and tunable axial
confinement. They therefore offer a pathway toward micrometer-scale
demonstrations in neutral-atom arrays. As a concrete reference platform, we
propose the spectroscopically established
Cs(\(6s\))--Cs(\(6s\))--Cs(\(nd_{3/2}\)) Rydberg trimer~\cite{Fey2019}. Its
near-degenerate vibrational \(E\)-doublet can serve as a candidate logical
qubit. In an optical-tweezer array, phase-controlled bond-length modulations
can steer the triangle through closed cycles on Kendall's shape sphere. These
cycles realize the Wilson loops of the shape-space connection. The resulting
non-Abelian holonomy can then be read out interferometrically through the trace
of the Wilczek--Zee Wilson loop. In the Appendices, we outline a concrete
control protocol together with platform parameters and operating conditions
that suppress leakage from the \(E\)-manifold.

More broadly, vibrational control structures are not limited to Rydberg
trimers. They can arise whenever the internal motion of a deformable triangle
supports a two-level system akin to a gapped, near-degenerate vibrational
\(E\)-doublet. Potential additional examples include Efimov-like trimers in
ultracold gases~\cite{Ferlaino-2010}. Thus, the shape-space connection
framework developed here may apply beyond Rydberg trimers to deformable
three-body systems with a suitable two-level computational subspace. In this
way, it points to a broad route toward geometric and topological control in
few-body quantum systems.

The outline of the article is as follows. We begin by reviewing the topology and geometry of Kendall's shape space for general
triangular structures~\cite{Kendall-1999}. Building on this framework, we construct the Wilczek--Zee connection that governs the
non-Abelian holonomy generated by adiabatic shape changes. We then identify the associated Wilson loop as the holonomic quantum
gate acting on a two-level subspace such as a vibrational $E$-doublet, and outline a perturbative scheme for evaluating its trace.
For arrays of triangular units, we show how entanglement between trimers can be encoded in multi-trace Wilson loops, and how the
resulting relative phases admit a Chern--Simons description for the Wilczek--Zee gauge field. As explicit examples, we construct a
$\pi/2$ phase gate and a Hadamard-type gate for a single triangle, and combine them with the Chern--Simons linkage mechanism to
outline a two-qubit entangling CNOT gate for two trimers. Finally, in the Appendices we first demonstrate how the Wilczek--Zee connection
evades Painlev\'e's theorem, which would otherwise preclude shape-space holonomy in time-reversal-symmetric settings. We then analyze
a Cs($6s$)--Cs($6s$)--Cs($nd_{3/2}$) Rydberg trimer in optical tweezers as a concrete candidate platform. The present work focuses on
geometric and holonomic considerations together with general estimates, and detailed Hamiltonian-based studies of specific platforms will
be presented elsewhere. 

%

\section{The Wilczek-Zee connection}
   
We start by  constructing the (adiabatic) SU(2)  Wilczek-Zee connection \cite{WilczekZee1984} for a generic three-body 
system.  We label with $\mathbf{r}_{\mathrm a}(t)$ ($\mathrm a=1,2,3$) the time-dependent positions of the three objects with masses $m_{\mathrm a}$
viewed as  the vertices of a (virtual) time-dependent triangle  
$\Delta(t)$, and we assume that the center of mass is at rest at the origin 
\[
m_1  \mathbf{r}_1 + m_2 \mathbf{r}_2 + m_3 \mathbf{r}_3 = 0.
\]
We invoke Guichardet’s theorem~\cite{Guichardet-1984}, which states 
that when the total orbital angular momentum vanishes, the  \(\mathbf{r}_{\mathrm a}(t)\) 
are classically  restricted to vibrations on the plane of the triangle. The geometric rotation induced by these vibrations is encoded in a 
$\mathrm{U}(1)$ connection $A$, which determines the holonomy in the corresponding shape space of triangles
\cite{Guichardet-1984, Iwai-1987, Shapere-1989b, Littlejohn-1997}.
Notably in the semiclassical Born–Oppenheimer regime this  $\mathrm{U(1)}$  connection manifests spectroscopically as a Berry phase 
producing an observable sign change of the electronic wavefunction over a 
closed Jahn–Teller vibrational cycle~\cite{Mead-1992,Yarkony-1996}.

Classically, Guichardet’s $U(1)$ connection $A$ encodes the geometric rotation
generated by planar vibrations of the triangle about its normal. But  in a quantum
three-body system with total angular momentum $L_{\rm tot}=0$ the ground state is
rotationally invariant. Consequently  the overall spatial orientation, including the
direction of the normal, is a gauge redundancy rather than a physical
observable. For this, after removing the center of mass the internal
configuration space is $\mathbb{R}^6\simeq\mathbb{R}^+\times\mathbb{S}^5$,
where $\mathbb{R}^+$ gives the size and $\mathbb{S}^5$ is the preshape sphere
of centered unit-size triangles. Quotienting by global rotations identifies preshapes 
related by $SO(3)$ and yields Kendall's shape sphere 
$\mathbb{S}^5/SO(3)\simeq\mathbb{S}^2_{K}$~\cite{Kendall-1999}. The resulting 
configuration manifold is $\mathbb{R}^+\times\mathbb{S}^2_{K}$,  with an overall size (breathing) 
coordinate along $\mathbb{R}^+$ and two independent
shape coordinates on $\mathbb{S}^2_{K}$. In many triangular molecular systems, the lowest 
excitations at fixed size form a gapped, near-degenerate vibrational (i.e. shape changing) 
$E$-doublet associated 
with these shape degrees of freedom. The adiabatic transport of the $E$-modes in shape space 
is governed by the non-Abelian Wilczek–Zee connection as an $SU(2)$ gauge field over
$\mathbb{S}^2_{K}$, and the associated kinematic frame bundle for closed
vibrational loops is the Hopf fibration $\mathbb{S}^3_{K}\to\mathbb{S}^2_{K}$
with $U(1)$ fiber.

We describe $\mathbb{R}^+ \times \mathbb{S}_K^3 \sim \mathbb{R}^+ \times \mathbb{S}_K^2 \times \mathbb 
S_K^1 $ by two complex Jacobi coordinates \((z_1, z_2)\)
and introduce  a  regular map $ Z = (Z_1(z_1, z_2),\, Z_2(z_1, z_2)) $ from $\mathbb{R}^+ \times \mathbb{S}_K^3$ onto
itself,  with coordinates 
\begin{equation}
Z \ = \ \left( \begin{matrix} Z_1 \\ Z_2 \end{matrix}\right) 
=
| Z |  \left( \begin{matrix} \cos\frac{\vartheta}{2} e^{i \phi_1} \\[-1.em]  \\  \sin\frac{\vartheta}{2} e^{i \phi_2}   \end{matrix}\right)  \ \ ; \ \ \ | Z |  =\sqrt{ |Z_1|^2 + |Z_2|^2},  
\la{z12}
\end{equation}
with 
\[
\phi =   \phi_2 -  \phi_1, 
\]  
identified as an internal coordinate and  
\[
\chi = - \frac{1}{2} (\phi_1+\phi_2) 
\]  
is external coordinate. Guichardet's $U(1)$ connection~\cite{Guichardet-1984,Iwai-1987,Shapere-1989b,Littlejohn-1997} 
that governs parallel transport on
Kendall's shape sphere via the Hopf fibration of the preshape three-sphere with circle fibers over $\mathbb{S}^2_K$,
can be written as
\begin{equation} 
A  \ \equiv  \ \frac{i}{2}   \frac{Z^\dagger \mathrm dZ - (\mathrm d Z^\dagger) Z}{ | Z |^2}  =   \mathrm d\chi  +   
 \frac{1}{2}  \cos \vartheta \mathrm d\phi.  
\la{dtheta2}
\end{equation}
Notably this  coincides with the connection of a single Dirac monopole.
The Bloch vector $\mathbf n(\theta,\phi)$ of the vibrational $E$–doublet is
a map from Kendall’s shape sphere $\mathbb S^{2}_{K}$ to the Bloch sphere and it
serves as the local quantization axis of the doublet. By a suitable, generally
local, gauge choice this axis may be aligned with the triangle’s body-frame
normal. With $\boldsymbol \sigma = (\sigma_x,\sigma_y,\sigma_z)$ the three Pauli matrices, the full $\mathrm{SU}(2)$
Wilczek–Zee connection then reads~\cite{Faddeev-1999,Duan-1979,Cho-1980,Kondo-2015}
\begin{equation}
\mathcal{A} =
\Big( A \,\mathbf n + \mathrm d\mathbf n \times \mathbf n \Big)\cdot\frac{\boldsymbol\sigma}{2i}
\;+\;
\Big( \rho\,\mathrm d\mathbf n + \eta\,\mathrm d\mathbf n \times \mathbf n \Big)\cdot\frac{\boldsymbol\sigma}{2i}.
\la{calA}
\end{equation}
Here $d\mathbf n$ and $\mathbf n\times d\mathbf n$ encode the geometric
variation of the Bloch vector over $\mathbb S^{2}_{K}$, and $(\rho,\eta)$
control the transverse twisting of the $E$–doublet basis as the shape moves on
$\mathbb S^{2}_{K}$. With 
\begin{equation}
\psi=\rho+i\eta,
\la{psi}
\end{equation}
 the pair $(A,\psi)$ forms an
Abelian Higgs multiplet under local ${U}(1)$ rotations about $\mathbf n$.
Together this gives six internal variables, as expected for
three points in $\mathbb{R}^3$ after removing the center of mass. Since
(\ref{calA}) transforms covariantly under such ${U}(1)$ rotations
\cite{Faddeev-1999}, any local phase generated by shape motion can be absorbed
into a gauge rotation about the quantization axis.

\section{Holonomic gates}

Continuous shape changes of the triangle $\Delta(t)$ induce parallel transport along 
a trajectory $\Gamma(t)$ on Kendall's shape space. In the present case, the relevant 
connection is the pullback of the Wilczek--Zee connection~(\ref{calA}) to $\Gamma(t)$. The
corresponding transport within the vibrational $E$-manifold that defines the qubit is 
described by the Wilson line $U_\Gamma(t)$, which solves
\begin{equation}
\frac{dU_\Gamma}{dt}=-\,\mathfrak q\,\mathcal A(t)\,U_\Gamma(t) \qquad U_\Gamma(0)=\mathbf 1,
\la{WilsonODE}
\end{equation}
with $\mathfrak q$ a charge and $\mathcal A(t)$ denotes the connection (\ref{calA}) 
evaluated along the trajectory. In particular, for a closed trajectory $\Gamma(0)=\Gamma(T)$
and with $\mathcal P$ denoting path ordering, the resulting holonomy is the endpoint value
\[
W_\Gamma \equiv U_\Gamma(T)
= \mathcal P \exp\!\left[-\mathfrak q\int_{0}^{T}\mathcal A(t)\,dt\right]
\]
\begin{equation}
= \mathcal P \exp\!\left(-\mathfrak q\oint_\Gamma \mathcal A\right).
\la{U}
\end{equation}
This holonomy is the unitary quantum gate acting on the logical subspace formed by the instantaneous 
two-fold degenerate $E$-manifold. By construction it is gauge covariant, while physical observables 
derived from it are gauge invariant and geometrically robust.
We also identify  $\mathfrak q$ as the effective Cartan weight in units where the minimal nonzero weight is $1/2$ due  to Dirac quantization,
that characterizes how the chosen two-level subspace couples to the diagonal (Cartan) component of the Wilczek--Zee connection.
The specific value of $\mathfrak q$ is fixed by the microscopic realization of the underlying $E$-doublet and can be determined by calibration.
With  generic ($A,\psi,\mathbf n$) and denoting $D=d+iA$,  the  curvature 
two-form of (\ref{calA}) is
\[
\mathcal F
=
\left(
d\mathcal A^{\mathrm c} 
+\epsilon^{\mathrm a \mathrm b \mathrm c}\,
\mathcal A^{\mathrm a}\wedge\mathcal A^{\mathrm b}
\right)
\frac{\sigma_{\mathrm c}}{2i}
\]
\[ 
= \Big[ d A + \tfrac12\big(|\psi|^2-1\big)\,\omega \Big]
\mathbf n \cdot \frac{\boldsymbol\sigma}{2i}  
 + \tfrac12 \Big[ D\psi \times \mathbf e^* + D^*\psi^*\times \mathbf e \Big]
\cdot \frac{\boldsymbol\sigma}{2i}
\]
\begin{equation}
d\omega =  \frac{1}{2}\, \mathbf n \cdot d \mathbf n \wedge d \mathbf n \  \ \ \& \ \ \   \mathbf e = d\mathbf n + i \mathbf n \times d\mathbf n.
\la{nots}
\end{equation}
It spans the entire $SU(2)$ Lie-algebra, so that by Ambrose-Singer theorem \cite{AmbroseSinger1953} 
the connection (\ref{calA}) has restricted holonomy group $SU(2)$. This ensures
that (\ref{U}) provides universal qubit control.

The adiabatic character of (\ref{U}) is therefore an assumption of the
present construction, not a claim of optimal gate speed. Existing
nonadiabatic holonomic gates remove this adiabatic requirement by
engineering cyclic subspace evolution directly. In the present work,
we instead use the adiabatic limit to expose the geometric gauge
structure of deformable triangular systems. Possible nonadiabatic or
shortcut implementations of the same shape-space holonomies are left
as a natural extension.

 To evaluate (\ref{U}) we introduce 
local orthonormal frames $\{ |n_\pm (t)\rangle\}$ along $\Gamma$
\begin{equation}
\mathbf n \cdot \boldsymbol \sigma |n_\pm\rangle = \pm |n_\pm\rangle \ \ \& \ \ \  
 |n_{+}\rangle \langle n_{+}|  + | n_{-}\rangle \langle n_{-}| =  \mathrm 1.
\label{n+-}
\end{equation}
We  discretize  $\Gamma$ into infinitesimal 
segments of duration $\Delta t$. Setting $t_k = k \Delta t$ and with
\[
U_{\alpha_{k+1}\alpha_k}(t_k)
=\big\langle n_{\alpha_{k+1}}(t_{k+1})\big|\,
e^{-q\,\mathcal A(t_k)\,\Delta t}\,
\big|n_{\alpha_k}(t_k)\big\rangle, 
\]
and with $\alpha_k=\pm$ labeling the instantaneous eigenbasis,
we write the path-ordered exponential (\ref{U}) as  the product
\begin{equation}
W_\Gamma
= \! \lim_{N\to\infty}\!
\sum_{\alpha_k=\pm} \!
|n_{\alpha_N}(t_N)\rangle \!
\prod_{k=0}^{N-1} \! U_{\alpha_{k+1}\alpha_k}(t_k)
\langle n_{\alpha_0}(t_0)|. 
\la{U2}
\end{equation}
As a product of $2\times 2$ matrices, this representation is particularly suited for actual computations.

The trace of the Wilson loop (\ref{U}) 
measures how much the qubit 
is twisted when transported once around a closed loop $\Gamma$. 
By gauge invariance this trace can only depend on 
the charged scalar $\psi = \rho + i\eta$ and the transverse one--form 
$\mathbf e$ introduced in (\ref{nots}) together with their complex conjugates, in gauge--covariant combinations.
In the instantaneous eigenbasis (\ref{n+-}) of $\mathbf n \cdot \boldsymbol\sigma$ we decompose  
the connection 
\begin{equation}
\mathcal A \ \buildrel{def}\over{=} \  C  \,  \frac{\sigma_z}{2i} + \mathcal A^{off} \ \equiv 
\ (A+\omega)  \frac{\sigma_z}{2i} + J\,\sigma_+ + J^*\sigma_-,
\la{fad-1}
\end{equation}
where $C= A + \omega$ is the diagonal abelian part, and $ \mathcal A^{off} $ denotes the off-diagonal part:
With spherical coordinates 
\begin{equation}
\mathbf n = \left( \begin{matrix} \cos\lambda \, \sin\mu \\ \sin\lambda\,\sin\mu \\ \cos\mu \end{matrix} \right),
\la{n-ap}
\end{equation}
we have \cite{Faddeev-1999b}
\[
dC = d A + \frac{1}{2}\, \mathbf n \cdot d \mathbf n \wedge d \mathbf n 
\]
\begin{equation}
= \frac{1}{2}\sin\vartheta d\phi \wedge d\vartheta
+ \sin\mu d\mu \wedge d\lambda ,
\la{fad-2} 
\end{equation}
and \[
\mathcal A^{off} = J\,\sigma_+ + J^*\sigma_- 
\]
\begin{equation}
= \psi\,\big(d\mu - i \sin\mu\, d\lambda\big) \sigma_+ + \psi^* \big(d\mu + i \sin\mu\, d\lambda\big) \sigma_-
\la{fad-3}
\end{equation}
for the off-diagonal part. The Wilson loop can then be evaluated  by treating the diagonal part exactly and 
expanding systematically in the transverse components using a 
time--ordered Dyson expansion. In this way 
we obtain from (\ref{U2}) in the continuum limit 
\begin{equation}
\mathrm{Tr}\,W_\Gamma
=
2\cos\!\left[\frac{\mathfrak q}{2}\oint_\Gamma C \right]
\Big( 1 -\mathcal I_2 +\mathcal I_4 +O(|\psi|^6)
\Big) .
\la{finalA}
\end{equation}
Here the $\mathcal I_2(t), \mathcal I_4(t), ... $ are gauge invariant quantities in powers of $J(t)$ and $J^*(t)$
that we construct iteratively in powers of $|\psi|^2$ 
as solutions of the Dyson equation \cite{Peskin-1995}
\[
F(t) = 1 - \frac{\mathfrak q^2}{4} \int\limits_0^t dt_1 \int\limits_0^{t_1}dt_2 J(t_1)e^{ \mathfrak q \int\limits_{t_2}^{t_1} C } J^* 
(t_2)  F(t_2).
\]
The result confirms  \(\psi\) as a geometric control knob for the holonomic gate. 

Remarkably, with $T$ the period of $\Gamma$ the non-vanishing  rotation angle $\Theta$ of the qubit 
\begin{equation}
\Theta(T) = 2 \arccos \left( \frac{1}{2} \mathrm{Tr}\,W_\Gamma \right)  = \mathfrak q \oint_\Gamma (A + \omega) + \mathcal O(|\psi|^2)
\la{bigthe}
\end{equation}
gives rise to an effective  geometric angular momentum contribution $\mathbb L_{\text{eff}}$ even in the absence of any
dynamical angular momentum, with magnitude
\begin{equation}
\mathbb L_{\text{eff}}  \ \approx   \ 2  \frac{ \mathbb I_{\Delta}}{T} \, 
\arccos \left( \frac{1}{2} \mathrm{Tr}\,W_\Gamma \right) .
 \la{eff} 
\end{equation}
Here  $ \mathbb I_{\Delta}$ is the average  moment of inertia of the three 
points $\mathbf r_{\mathrm a}(t)$ around the average axis of the central angle,
over the period $T$.

Finally, in extended architectures we  encounter multiple triangular units such as 
arrays of Rydberg trimers in optical-tweezer geometries. In that setting,
entanglement between distinct triangular qubits becomes essential. The three-dimensional
Chern--Simons functional provides a compact, gauge-invariant way to encode this entanglement.
For the gauge field~(\ref{calA}), and up to an exact three-form, it takes the form
\[
{\rm ChS}[\mathcal A] 
= \frac{k}{4\pi} \int \mathrm{Tr}\!\left(
  \mathcal A \wedge d\mathcal A 
  + \tfrac{2}{3}\, \mathcal A \wedge \mathcal A \wedge \mathcal A 
\right) 
\]
\vspace{-1.7em}
\begin{equation*}
= - \frac{k}{4\pi} \int \left\{
  \tfrac12\, C \wedge dC
  - i \big[ \psi^* D\psi - \psi\, \bar D \psi^* \big] 
    \wedge \omega 
\right\},
\end{equation*}
where $k\in\mathbb Z$ denotes the level, and the integral is over the pullback of $\mathbb S^3_K$ defined by the
time-periodic triangle worldvolume. On the $\psi=0$ truncation, the averaged
multi-trace Wilson loop for several triangular structures yields~\cite{Polyakov-1988,Witten-1989}
\[
\int\![dC]\,
 \exp\Big\{ i \frac{k}{8\pi}\!\int C\wedge dC
     + i\sum_i \mathfrak q_i\!\oint_{\Gamma_i}\! C \Big\}
\]
\vspace{-1.7em}
\begin{equation}
= \exp\!\left\{\! 
   \frac{4\pi i}{k}\!\sum_{i\neq j}\! \mathfrak q_i \mathfrak q_j\,\mathrm{Lk}(\Gamma_i,\Gamma_j)
 + \frac{2\pi i}{k}\!\sum_i \mathfrak q_i^2\,{\rm SLk}(\Gamma_i)\!
 \right\},
\la{link}
\end{equation}
where the linking ($\mathrm{Lk}$)  and self-linking (${\rm SLk}$)  quantify the topological contribution to two-qubit entangling phases.
The result  mirrors entanglement of Abelian anyons with charges $\mathfrak q_i$ and
statistics $4\pi \mathfrak q_i \mathfrak q_j/k$~\cite{Nayak2008TQC}.

\section{examples of gates}

\subsection{Single qubit gates}

As concrete examples of the general theory, we proceed to demonstrate how to
construct loops on Kendall's shape sphere \(\mathbb S^2_K\) that realize a
\(\pi/2\) phase rotation gate and a Hadamard-type gate, respectively, on the
\(E\)-doublet. To this end, we take the equilibrium trimer's initial shape to
have polar coordinates \((\theta_0,\phi_0)\) on \(\mathbb S^2_K\) and introduce
a small elliptical loop
\begin{equation}
\Gamma:\quad \left\{ 
\begin{matrix}  \hspace{-0.45cm} \theta(s) = \theta_0 + a\cos s
\\[2pt]
\hspace{0.1cm}  \phi(s) = \phi_0 + \dfrac{b}{\sin\theta_0}\sin s
\end{matrix} \right.
\qquad
s\in[0,2\pi),
\la{ellipse}
\end{equation}
with \(a,b\ll 1\) setting the loop size. To leading order in \(a,b\), this loop
encloses the solid angle
\begin{equation}
\Omega_\Gamma \simeq \pi a b .
\la{ellipse2}
\end{equation}
For the computational basis, we choose the \(E\)-doublet eigenstates at the
base point,
\begin{equation}
|0\rangle \equiv |E^{(1)}\rangle \ \ \ \ \& \ \ \ \ 
|1\rangle \equiv |E^{(2)}\rangle.
\la{basis}
\end{equation}
This choice fixes a frame for the degenerate \(E\)-manifold at
\((\theta_0,\phi_0)\). In using it to describe the evolution around
\(\Gamma\), we assume the adiabatic approximation: the state remains within the
instantaneous \(E\)-doublet along the loop, and the resulting action on the
computational basis is given by the Wilczek--Zee holonomy.

\subsubsection{$\pi/2$ phase gate about the $z$ axis}

We work in the pinned--normal regime, where the normal vector $\mathbf{n}$ is locked to the bias field. 
In this case, the leading-order contribution in $|\psi|$ to the Wilczek--Zee connection coincides with the 
Guichardet connection along a fixed Pauli axis, which we take to be $\sigma_z$ in the basis~(\ref{basis}).
The holonomy  (\ref{U}) for the loop $\Gamma$ becomes
\begin{equation*}
W_\Gamma
= \mathcal{P}\exp\!\left( -  q\oint_\Gamma \mathcal{A} \right)
\simeq
\exp\!\left(-\frac{i}{2}\,\Theta_\Gamma\,\sigma_z \right) .
\end{equation*}
Here
\[ 
\Theta_\Gamma = \frac{q}{2}\,\Omega_\Gamma + \mathcal{O}(|\psi|^2),
\]
with $q$ the holonomy charge, in line with the general relation (\ref{finalA})
for the Wilson–loop trace.
The resulting gate is the single–qubit phase rotation
\[
U_\Gamma \equiv W_\Gamma
\simeq
\exp\!\left(-\frac{i}{2}\,\Theta_\Gamma\,\sigma_z \right)
=
\begin{pmatrix}
e^{-i\Theta_\Gamma/2} & 0 \\
0 & e^{+i\Theta_\Gamma/2}
\end{pmatrix}.
\]
To obtain a $\pi/2$ rotation, we choose the loop area such that
\[
\Theta_\Gamma = \frac{\pi}{2}
\quad\Longrightarrow\quad
\Omega_\Gamma = \frac{\pi}{q}
\ \Rightarrow \ 
a b = \frac{1}{q}.
\]
With this choice the holonomy takes the explicit form
\begin{equation}
U_{\Gamma}^{(\pi/2)}
=
\exp\!\left(-\frac{i}{4}\pi\,\sigma_z\right)
=
\frac{1}{\sqrt{2}}
\begin{pmatrix}
1 - i & 0 \\
0 & 1 + i
\end{pmatrix},
\la{pi2}
\end{equation}
which is the standard $\pi/2$ phase gate about the $z$ axis.

\subsubsection{Hadamard gate from a rotation about the $y$ axis}

For a Hadamard-type gate we require a loop whose holonomy is a $\pi/2$ rotation
about an axis in the equatorial plane of the Bloch sphere. In the representation
(3) of the Wilczek--Zee connection, and after an appropriate gauge
choice and rescaling of $\psi$, we write along the loop $\Gamma_H$
\[
\mathcal A(s) =
C(s)\,\frac{\sigma_z}{2i}
\;+\;
\mathcal A_\perp(s),
\]
with 
\[
\mathcal A_\perp(s)
=
\mathrm{Re}\,\psi(s)\,\frac{\sigma_x}{2i}
\;+\;
\mathrm{Im}\,\psi(s)\,\frac{\sigma_y}{2i},
\]
where $s\in[0,2\pi)$ parametrizes the loop $\Gamma_H$. 
Solving for the Wilson line $U(s)$ from (\ref{WilsonODE}) 
we obtain  for  the holonomy
\begin{equation}
W_{\Gamma_H} \equiv U(2\pi)
= \mathcal P \exp\!\left[- q\int_0^{2\pi} \mathcal A(s)\,ds\right].
\end{equation}

 We start with the Abelian evolution
\[
U_z(s) = \mathcal P \exp\!\left[-q\int_0^s
C(s')\,\frac{\sigma_z}{2i}\,ds'\right],
\]
\[
=
\exp\!\left[- q \,\frac{\sigma_z}{2i} \int_0^s C(s')\,ds'\right],
\]
since the path ordering is redundant. 
We then write
\[
U(s) = U_z(s)\,V(s).
\]
and substitute this into (\ref{WilsonODE}),
\[
\frac{d}{ds}(U_z V)
=
\left(\frac{dU_z}{ds}\right)V + U_z\frac{dV}{ds}
\]
\[
=
-  q\left(C \frac{\sigma_z}{2i} + \mathcal A_\perp\right)U_z V.
\]
Since $U_z$ itself satisfies
\begin{equation}
\frac{dU_z}{ds} = -\, q\,C(s)\,\frac{\sigma_z}{2i}\,U_z(s),
\end{equation}
the diagonal terms cancel and we are left with
\begin{equation}
U_z\,\frac{dV}{ds} = -\, q\,\mathcal A_\perp\,U_z\,V.
\end{equation}
Multiplying from the left by $U_z^{-1}(s)$ yields the interaction--picture
equation
\[
\frac{dV}{ds} = -\,q\,\mathcal A_\perp^{(I)}(s)\,V(s) \ \ \ \ \  V(0)=\mathbf 1,
\]
where we have defined
\begin{equation}
\mathcal A_\perp^{(I)}(s) := U_z^{-1}(s)\,\mathcal A_\perp(s)\,U_z(s).
\la{AperpI}
\end{equation}
The solution is
\[
V(2\pi)
=
\mathcal P\exp\!\left[- q\int_0^{2\pi} \mathcal A_\perp^{(I)}(s)\,ds\right]
\]
\[
= \mathcal P\exp\!\left[- q\int_0^{2\pi} U_z^{-1}(s)\,\mathcal A_\perp(s)\,U_z(s)\,ds\right].
\]
Thus we have the factorization
\[
W_{\Gamma_H} = U(2\pi) = U_z(2\pi)\,V(2\pi).
\]

Since $U_z(s)$ is a rotation around $\sigma_z$, the transformed transverse
generator $\mathcal A_\perp^{(I)}(s)$ remains in the $x$--$y$ plane and can be
written as
\[
\mathcal A_\perp^{(I)}(s)
=
|\psi(s)|\,
\big[
\cos\Theta(s)\,\frac{\sigma_x}{2i}
+
\sin\Theta(s)\,\frac{\sigma_y}{2i}
\big],
\]
where the angle $\Theta(s)$ depends on the phase of $\psi(s)$ and on the
accumulated $z$--phase 
\[
r(s)=  q\int_0^s A(s')\,ds'.
\]
By steering the control
phase of $\psi(s)$ along the loop, we can impose
\begin{equation}
\Theta(s) = \frac{\pi}{2}
\qquad\text{for all } s\in[0,2\pi),
\end{equation}
so that in the interaction picture the non-Abelian part of the connection is
aligned along $\sigma_y$ at all parameter values:
\begin{equation}
\mathcal A_\perp^{(I)}(s)
=
|\psi(s)|\,\frac{\sigma_y}{2i}.
\end{equation}
In the small-loop limit, where the curvature is approximately constant over the
enclosed area and higher-order path-ordering effects can be neglected, the
interaction--picture holonomy becomes
\begin{equation}
V(2\pi)
\simeq
\exp\!\left(
-\,i\,\frac{q}{2}\,\Omega_H\,\sigma_y
\right)
+
\mathcal{O}(\Omega_H^2,|\psi|^2),
\end{equation}
where $\Omega_H$ is the oriented area (solid angle) associated with the loop
in control space; for the elliptical loop~(\ref{ellipse}) it is given by
(\ref{ellipse2}). Writing
\[
V(2\pi)
\simeq
\exp\!\left(
-\frac{i}{2}\,\Theta_H\,\sigma_y
\right)
\]
and 
\[
\Theta_H
=
\frac{ q}{2}\,\Omega_H
+
\mathcal{O}(\Omega_H^2,|\psi|^2),
\]
a Hadamard-type transformation corresponds to a rotation by $\Theta_H=\pi/2$,
so we choose
\begin{equation}
\Theta_H = \frac{\pi}{2}
\quad\Longrightarrow\quad
\Omega_H = \frac{\pi}{q}.
\la{Omega-H}
\end{equation}
The associated non-Abelian gate is then
\begin{equation}
U_H
= 
\exp\!\left(-\frac{i\pi}{4}\sigma_y\right)
=
\frac{1}{\sqrt{2}}
\begin{pmatrix}
1 & -1 \\
1 & \phantom{-}1
\end{pmatrix},
\la{Hadam}
\end{equation}
which equals the standard Hadamard gate composed on the right with a Pauli
$\sigma_z$ and is therefore locally equivalent to the Hadamard gate in the
computational basis~(\ref{basis}). The full holonomy is
\[
W_{\Gamma_H}=U_z(2\pi)\,U_H,
\] 
where the Abelian factor $U_z(2\pi)$ is a
$z$--phase gate that can be calibrated or compensated using the phase--gate
protocol of the previous subsection.

The corresponding loop $\Gamma_H$ can be chosen to have the form~(\ref{ellipse}),
with $(a,b)$ selected to satisfy~(\ref{Omega-H}). Using~\eqref{ellipse2},
the Hadamard condition fixes
\[
a b = \frac{1}{ q}.
\]
Thus the $\pi/2$ phase gate~(\ref{pi2}) and the
Hadamard-type gate~(\ref{Hadam}) can both be implemented by loops with the
same enclosed solid angle $\Omega = \pi/{ q}$, but with different control of the
phase of $\psi(s)$: The phase-gate protocol of the previous subsection yields a
pure $z$--rotation, whereas enforcing $\Theta(s)=\pi/2$ along the
loop yields the Hadamard-type $y$--rotation $U_H$.

\subsection{Two-qubit holonomic CNOT gate}

We proceed to outline how (\ref{link}) 
can be used to build a two-qubit entangling gate between two
trimers, and how in combination with the above single-qubit gates this yields a CNOT gate.
For this we  consider two triangular structures labelled $u$ and $v$, each encoding a
logical qubit~(\ref{basis}) in its vibrational $E$--doublet. For a two-qubit
gate we assume that each triangle couples to the Abelian connection $C$ with a holonomy
charge that depends on its logical state:
\begin{equation}
Q_u =
\begin{cases}
0, & |0\rangle_u\\
Q, & |1\rangle_u
\end{cases}
\ \ \ \ \ {\rm and} \ \ \ \ \ 
Q_v =
\begin{cases}
0, & |0\rangle_v\\
Q, & |1\rangle_v
\end{cases}.
\la{qAB}
\end{equation}
Physically, this corresponds to a coupling of the Abelian part of the
Wilczek--Zee connection only to one branch of the $E$--doublet, e.g.\ via a
state-dependent light shift or auxiliary level, while the other branch is
effectively neutral. Here the charges $Q_u,Q_v$ should be viewed as 
platform-dependent but quantized Cartan weights assigned to the logical
states of the $i$th ($i=u,v$) trimer during the gate cycle. in particular, one branch 
can be engineered to be effectively neutral ($Q_i=0$) while the
other carries charge $Q_i=Q$.
For the gate cycle, we drive both trimers simultaneously through loops
$\Gamma_u$ and $\Gamma_v$ in their respective shape spaces, such that on the
shape space
\[
\mathrm{Lk}(\Gamma_u,\Gamma_v)=1
\ \ \ \ \ \& \ \ \ \ \ 
{\rm SLk}(\Gamma_u)={\rm SLk}(\Gamma_v)=0.
\]
For each computational basis state $|\alpha\beta\rangle$ the charges
$(Q_u,Q_v)$ then take the eigenvalues given by (\ref{qAB}), so that
the Chern--Simons path integral (\ref{link}) contributes a state-dependent
phase
\[
\exp\{i\Phi_{\alpha\beta}\}  = \exp\!\left\{ i \frac{4\pi}{k}\,Q_u Q_v\right\}.
\]
This defines a diagonal unitary matrix $U_{\mathrm{CS}}$ that acts on the
computational basis with eigenvalues given by these phases. Specifically, on
the computational basis $\{|00\rangle,|01\rangle,|10\rangle,|11\rangle\}$ we
have
\begin{equation}
U_{\mathrm{CS}}
=
\mathrm{diag}\!\left(
1,\;
1,\;
1,\;
e^{i\phi}
\right)
\ \ \ {\rm with} \ \ \ 
\phi
=
\frac{4\pi Q^2}{k}.
\la{U-CS}
\end{equation}
Choosing $k$ such that
\[
\phi = \pi
\quad\Longleftrightarrow\quad
\frac{4\pi Q^2}{k} = \pi
\;\Rightarrow\;
k = 4 Q^2,
\]
we obtain the standard controlled-$Z$ gate
\[
U_{\mathrm{CZ}}
\equiv
U_{\mathrm{CS}}(\phi=\pi)
=
\mathrm{diag}\!\left(1,1,1,-1\right)
\]
\begin{equation}
=
|0\rangle\!\langle 0|_u\otimes\mathbbm{1}_v
+
|1\rangle\!\langle 1|_u\otimes \sigma_z^v. 
\la{CZ}
\end{equation}
Any additional single-qubit phases accumulated during the joint loop can be
absorbed into the local phase gates constructed in the previous subsection.

Let now the first trimer $u$ be the control and the second trimer $v$ the target.
We denote by $U_H$ the holonomic Hadamard-type gate (\ref{Hadam}) acting on
$v$. Since it differs from the standard Hadamard gate by the action of
Pauli-$\sigma_z$ on the right, the canonical Hadamard on $v$ is
\[
H_v = U_H\,\sigma_z^v, 
\]
and on the two-qubit Hilbert space the corresponding operator is
$\mathbbm{1}_A\otimes H_v$.
The composite sequence
\[
U_{\mathrm{CNOT}}
=
(\mathbbm{1}_u\otimes H_v)\,
U_{\mathrm{CZ}}\,
(\mathbbm{1}_u\otimes H_v^\dagger)
\]
realizes a CNOT gate with $u$ as control and $v$ as target. Indeed,
\[
H_v\,\sigma_z^v\,H_v^\dagger = \sigma_x^v \equiv X_v,
\]
so that using \eqref{CZ} we obtain
\[
U_{\mathrm{CNOT}}
=
|0\rangle\!\langle 0|_u\otimes\mathbbm{1}_v
+
|1\rangle\!\langle 1|_u\otimes X_v .
\]

\noindent
Operationally, the CNOT is then obtained by:

\vskip 0.2cm
\noindent
(i) applying the holonomic Hadamard-type gate $H_v$ to the $E$--doublet of
trimer $v$,

\vskip 0.2cm
\noindent
(ii) executing a joint gate cycle in which trimers $u$ and $v$ traverse
linked loops $(\Gamma_u,\Gamma_v)$ in configuration space, producing the
Chern--Simons controlled phase (\ref{U-CS}), and

\vskip 0.2cm
\noindent
(iii) applying the holonomic gate $H_v^\dagger$ on $v$ (e.g.\ by running the
loop corresponding to $H_v$ in reverse).

\vskip 0.2cm
In combination with the single-qubit gates above, we may
now  construct a universal set of holonomic gates for arrays of
triangular qubits.

\section{Outlook:}

We have established the concept of shape space as a natural and experimentally accessible control 
manifold for non-Abelian holonomies in deformable three-body systems. For platforms whose 
low-energy vibrational spectrum contains a gapped, near-degenerate $E$-doublet, we have 
derived the Wilczek--Zee connection governing adiabatic transport on Kendall's shape sphere of triangles. 
The resulting restricted holonomy group is $\mathrm{SU}(2)$, so that closed loops in shape space 
implement universal single-qubit holonomic gates within the $E$-manifold. Moreover, the gauge-invariant 
trace of the associated Wilson loop provides a compact and experimentally accessible diagnostic of 
these gates, offering a direct signature of non-Abelian geometric transport.

Among the future theoretical challenges is the development of platform-specific mappings from explicit
microscopic trimer Hamiltonians to the effective gauge data. This  includes dynamical control of the 
knobs $(\mathbf{n},\psi)$ that enter the Wilczek--Zee connection and a quantitative characterization 
of the resulting control landscape. To proceed from few-body molecular structure to robust 
geometric gates, a systematic treatment of imperfections is also needed. This includes effects of symmetry breaking 
within the $E$-manifold, stray-field shifts, trap anisotropies, and coupling to spectator vibrational modes. 
Such detailed platform-specific investigations will clarify the parameter regimes in which shape-space 
holonomies provide a practical advantage for quantum control.

Beyond a single qubit, we have also proposed holonomic control in trimer arrays as a theoretically and experimentally 
challenging direction. The entangling mechanism based on linked holonomic cycles in shape space provides a framework for 
exploring multi-trimer architectures with two-qubit phases governed primarily by linking data. This  suggests  a 
pathway toward more  elaborate topological control primitives in programmable molecular arrays.

As a future experimental objective,  we propose a proof-of-principle demonstration  
on a bond-length--modulated Cs($6s$)--Cs($6s$)--Cs($nd_{3/2}$) 
Rydberg trimer trapped in optical tweezers. With details given in Appendix,  
we propose to (i) prepare and spectroscopically 
characterize an $E$-doublet in a single triangular unit, (ii) implement phase-controlled bond-length modulations 
that execute calibrated closed cycles on Kendall's shape sphere, and (iii) read out the resulting non-Abelian 
holonomy interferometrically using a Ramsey--echo protocol that refocuses dynamical phases while preserving 
the geometric signal. By varying the loop area, orientation, and repetition number, the measured 
Wilson-loop trace can be benchmarked against its predicted dependence on the control cycle, enabling 
a quantitative and gauge-invariant characterization of the effective Wilczek--Zee connection. In addition, we propose
the design of nonadiabatic holonomies as a potential route to faster gates without compromising the underlying geometric structure.

\vskip 0.2cm

\section*{Acknowledgements:}

 A.J.N. thanks  X.-C. Yao for discussions on molecular trimers, F. Wilczek for discussions on Painlev\'e's theorem, 
and R. Jaffe for critical comments on an early version.   J.D. and A.J.N thank X. Peng for a discussion.
A.J.N. is supported by the Swedish Research Council under Contract No. 2022-04037
and by COST Action CA23134 (POLYTOPO).

\section*{Appendices:}  

\subsection{Painl\'eve's theorem}

The Schr\"odinger equation governing  a trimer as an isolated, conservative many-body system is time-reversal invariant, provided 
magnetic fields are absent and weak interaction effects  are ignored.   In that case,  the classical ``falling cat" 
theorem by Painlev\'e~\cite{Painleve} 
states that  

\vskip 0.2cm
\noindent
{\it if the relative motion of an isolated system of particles is governed solely by conservative forces, and if at some instant 
$t=t_0$ all particles are at rest, then the system can never return to the same internal configuration of relative positions 
while having a different overall orientation in space.}

\vskip 0.2cm

This suggests that in our physical scenarios the curvature (\ref{fad-2}) 
of the Guichardet  connection  $A$ should in fact identically
vanish. 
Nevertheless, we now demonstrate 
through a broadly applicable example that, although Painlev\'e's theorem is correct,  its assumptions are not generic. 
For this 
we consider a (semi)classical trimer with bond  lengths   
\[
r_{\mathrm a \mathrm b }(t) =  |\mathbf r_{\mathrm a}- \mathbf r_{\mathrm b}|
\]  
that vibrate  around their  
time-averaged equilibrium values 
\[
d_{\mathrm a \mathrm b}\sim \, <\! r_{\mathrm a \mathrm b}(t)\! >.
\]
We Taylor expand the potential 
\[
V(r_{\mathrm a \mathrm b}) \approx V(d_{\mathrm a \mathrm b}) +{ \frac{1}{2}}V''(d_{\mathrm a \mathrm b}) (r_{\mathrm a \mathrm b}- d_{\mathrm a \mathrm b})^2 + ...,
\]  
where we keep only the leading, harmonic contribution. 
For a time reversal  invariant  trajectory,  without loss of generality, 
we allow bond $r_{12}$ to oscillate according to
\begin{equation}
r_{12}(t)  = d_{12}  +  \epsilon_{12} \cos(\Omega_{12} t), \ \ \ \ \ (\epsilon_{12} \ll \ d_{12})
\la{D12}
\end{equation} 
and we choose the vertex $3$ to  describe the Rydberg atom of the trimer. 
Since  the Rydberg excitation breaks the equilateral $D_{3h}$ symmetry into the 
$C_{2v}$ symmetry of an isosceles configuration, for the other two bonds we may choose
\mbox{$d_{13} \, = \, $}\mbox{$ \, d_{23}   = d $} 
and  $\epsilon_{13} \, = \, \epsilon_{23} \, = \, \epsilon $ (with $\epsilon \ll  d$) and
$\Omega_{13} \, =\, \Omega_{23}$  = $ \Omega $ so that
\begin{equation}
\begin{pmatrix} r_{13} (t)  \\  r_{23} (t) \end{pmatrix} = 
\begin{pmatrix} 
d  + \epsilon \cdot \cos(\Omega t + \phi_{13}  )   \\
d  + \epsilon  \cdot \cos(\Omega t + \phi _{23}  )  
\end{pmatrix},
 \la{nd}
\end{equation}
and in general the  bond $r_{12}(t)$ differs from $r_{13}(t)$ and $r_{23}(t)$ in its mean length, amplitude, and frequency. 
For generic \(\phi_{13}\) and \(\phi_{23}\) time reversal is then not a symmetry of the given trajectory (\ref{nd}). 
However,  it becomes a
symmetry  if \(\phi_{13} = -\phi_{23}\) and $t\to -t$ is combined 
with multiplication of (\ref{nd}) by Pauli matrix $i\sigma_y$ exchanging  the vertices 1 and 2. 
%
%
%
%
%
\begin{figure}
\centerline{\includegraphics[width=8cm]{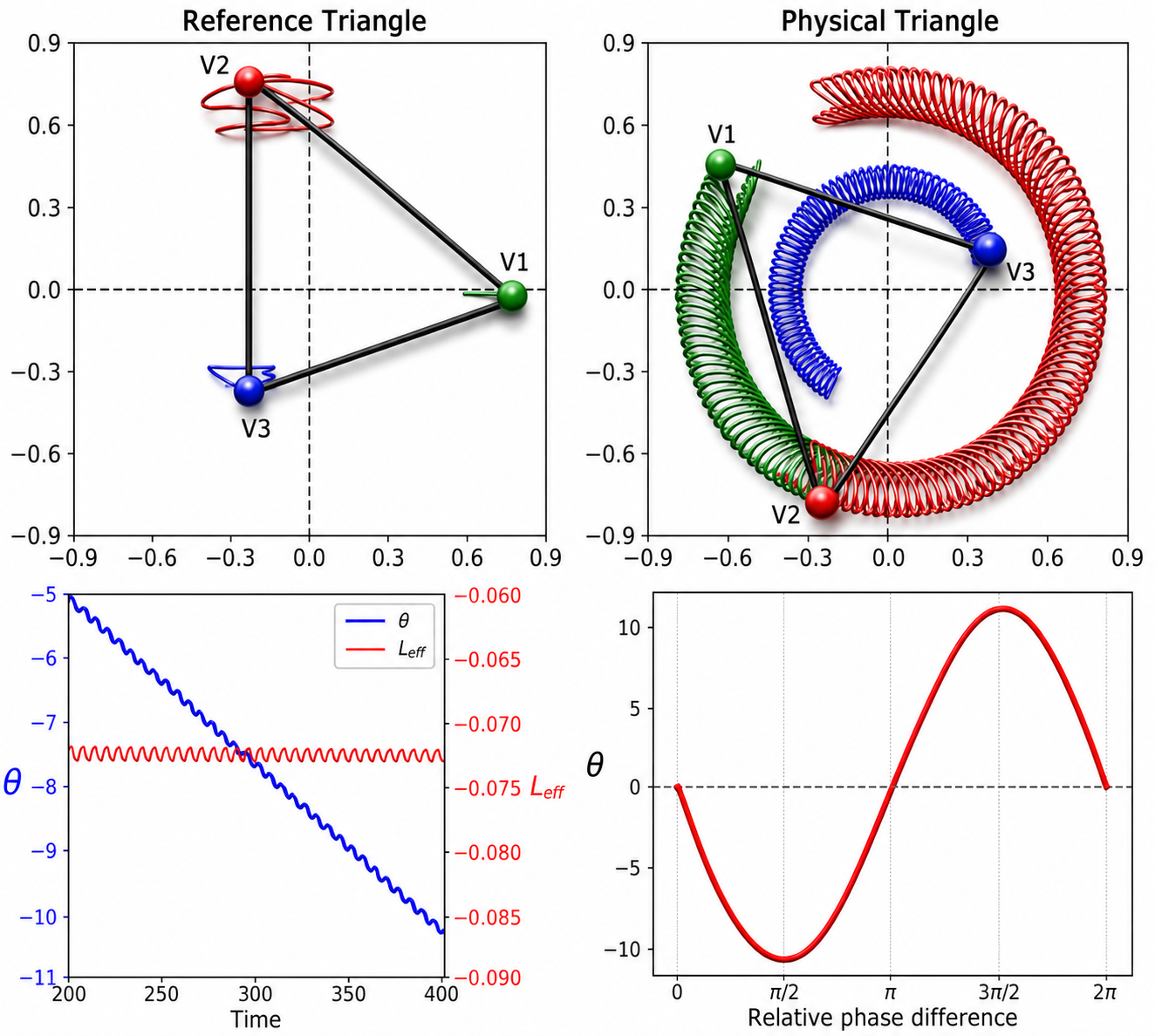}}
\caption{Panel a) A  representative  shape space time evolution of triangle $\triangle(t)$ with 
numerical parameter values $ d=1.0 \, \& \,   \epsilon=0.15 \, \& \,  d_{12} = 1.1  \, \& \,   \epsilon_{12} = 0.2   \, \& \,  \Omega / \Omega_{12} = 3  \, \& \, 
m_{1,2} = 2.1  \, \& \, m_3=4.7$.  
Panel b) Corresponding rotational motion in the physical space. Panel c)  Time evolutions of the angle $\theta(t)$   
and the effective angular momentum (\ref{eff}).
 Panel d) Comparison of  the rotational distance
(effective angular velocity) with different values of $\phi_{13} - \phi_{23}$. 
}
\label{fig-1}
\end{figure}
%
%
%
%
%
%
%
We have performed numerical investigations with  (\ref{D12}),  (\ref{nd}) to confirm that, despite Painlev\'e's theorem, 
for a  generic  time reversal symmetric trajectory 
 the vibrational modes induce rotational motion of the triangle  around its normal axis, even with zero mechanical
angular momentum. Thus the  Guichardet connection $A$ and in particular the geometric angular momentum
(\ref{eff}) can not vanish.  Figure~\ref{fig-1} illustrates this 
in both shape space (panel a) and physical space (panel b). Panel c shows the time evolution of both the rotation 
angle $\theta(t)$ and the effective angular momentum (\ref{eff}) 
over an extended simulation. 
As the ratio of the sampling time-step to the trajectory length decreases, $\theta(t)$ exhibits linear growth in time
while the geometric angular momentum $\mathbb{L}_{\text{eff}}(t)$ in (\ref{eff}) 
approaches a constant value, implying that the motion asymptotically 
resembles uniform rotation of a rigid trimer, despite the absence of mechanical angular momentum.  
Finally, panel d demonstrates that for generic \(\phi_{13} = -\phi_{23}\)
rotational motion is present, and vanishes only when 
$\phi_{13} = \phi_{23} = 0 \ \text{mod} \, \pi$. This is also the sole value where all three bond lengths come to a stop
simultaneously, fully in line with Painlev\'e's theorem. 
Notably,  the effective angular velocity  reaches its maximum value at 
$ \phi_{13} = -\phi_{23} = \frac{\pi}{4} + \frac{k\pi}{2}$ with  $k \in \mathbb{Z} $. 
This is also  the phase value that emerges from the solution to the equations of motion derived from the Lagrangian
\begin{equation*}
\mathcal  L = \frac{1}{2} \left( \dot{r}_{13} r_{23} - \dot{r}_{23} r_{13} \right) - \frac{ \Omega}{2} \left[ (r_{13} - d)^2 + (r_{23} - d)^2 \right],
\end{equation*}
describing uniform circular precession around a fixed point. The  enclosed phase-space area over a
cycle generates a Berry phase of $\pi$ which  is characteristic of a spin-$\frac{1}{2}$ system \cite{Alekseev-1988} that
has been observed in molecular triangles~\cite{Mead-1992,Yarkony-1996}.

\subsection{Application to Cesium trimer}

As a demonstrator we propose a Cs($6s$)--Cs($6s$)--Cs($nd_{3/2}$)
Rydberg trimer confined within a single site of an optical tweezer array. We
specify a concrete operating point and control protocol that realize a
vibrational qubit encoded in the near-degenerate $E$–doublet, together with an
experimentally accessible Wilson-loop trace of the corresponding
Wilczek--Zee connection.

\subsubsection{Setup:} 

Spectroscopic studies have established that triatomic
Cs($6s$)--Cs($6s$)--Cs($nd_{3/2}$) Rydberg molecules can be formed by binding
one excited Cs atom to two ground-state atoms~\cite{Fey2019}, with
characteristic bond lengths $R_{1}=R_{2}\!\sim\!2000\,a_{0}$
($\simeq 0.1~\mu\mathrm{m}$). Typical optical tweezers provide confinement on
the micrometer scale and can be arranged in programmable
arrays~\cite{SaffmanRMP2010,BrowaeysLahaye2020}, allowing the entire trimer to
fit comfortably within a single trap volume. In this configuration the two
perturbers remain bound within the Rydberg electron’s molecular potential,
while the tweezer supplies overall center-of-mass confinement; see Figure 2.
%
%
%
%
%
\begin{figure}
\centerline{\includegraphics[width=8cm]{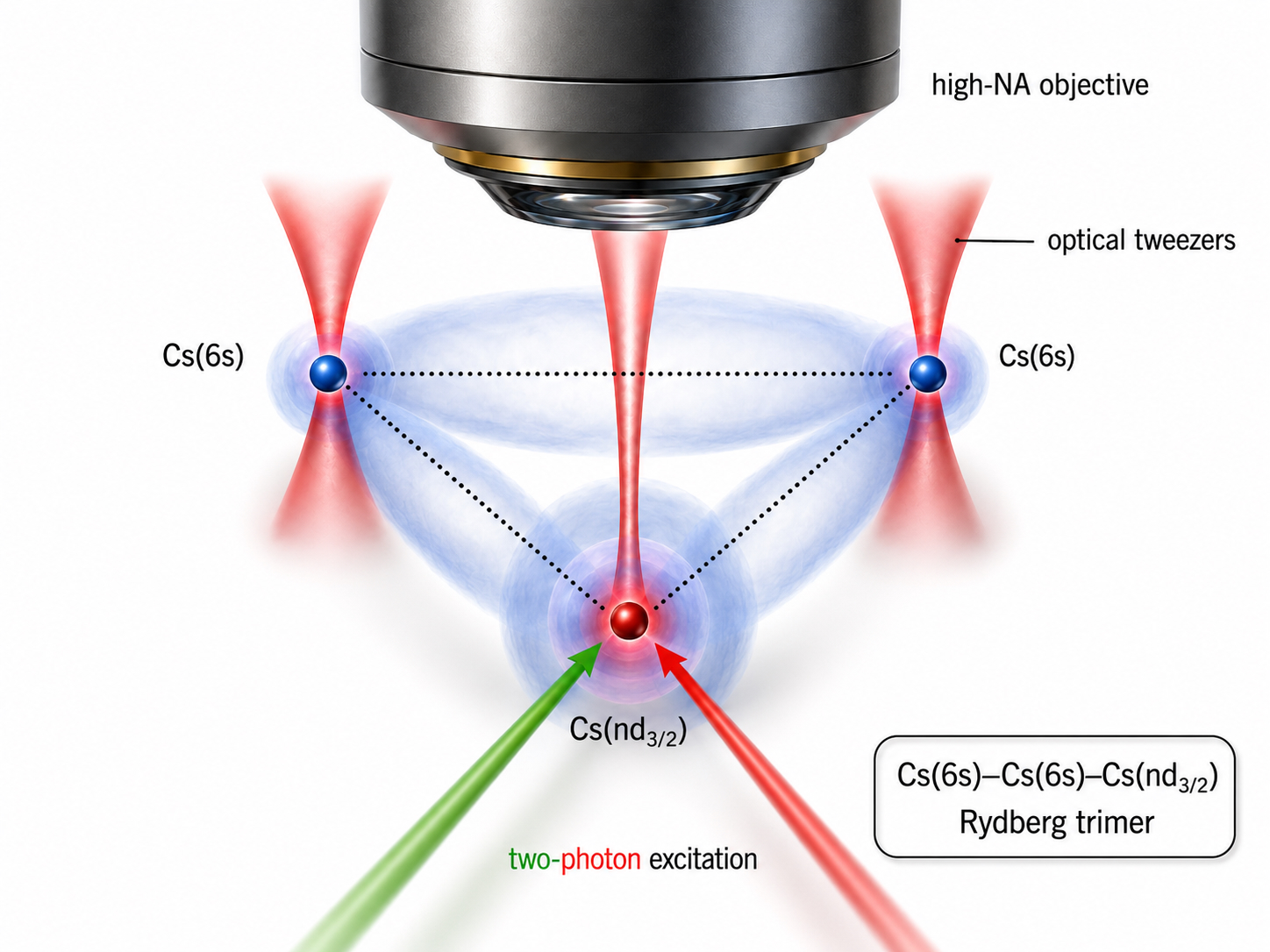}}
\caption{\textbf{Outline of proposed  Cs Rydberg trimer in optical tweezers.}
Three optical tweezers, focused by a high-NA objective lense, trap three Cs atoms in a triangular configuration,
realizing a Cs($6s$)--Cs($6s$)--Cs($nd_{3/2}$) Rydberg trimer.
Rydberg excitation is illustrated schematically by a two-photon beam pair addressing the central site. }
\label{fig-2}
\end{figure}
%
%
%
%
%
%
%
The lifetime of
a room-temperature Cs Rydberg atom is typically a few to a few tens of
microseconds increasing significantly under cryogenic
conditions~\cite{SaffmanRMP2010}. This timescale is compatible with the loop
durations required for shape-space control in the present
geometry~\cite{Fey2019,SaffmanRMP2010}. Figure \ref{fig-1}  illustrates
representative kinematics within a classical vibrational model.

The planar tweezer potential is used to tune the trimer’s apex angle such that
the equilibrium geometry lies close to $D_{3h}$ symmetry. In this set-up, the
lowest internal vibrational modes decompose into a breathing $A$-mode with
energy $\mathcal E_{A}$ and a near-degenerate pair of angular modes with energies
$\mathcal E_{E^{(1)}}$ and $\mathcal E_{E^{(2)}}$ that span the logical $E$-manifold. Defining
the mean $E$-manifold energy
\[
\mathcal E_{E}\equiv \frac{\mathcal E_{E^{(1)}}+\mathcal E_{E^{(2)}}}{2},
\]
and denoting by $T_{\mathrm{loop}}$ the duration of a single holonomic loop in
shape space, our target is a near-degenerate but gapped window
\begin{equation}
\Delta_{\mathrm{gap}} \equiv \mathcal E_{A}-\mathcal E_{E} \;\gg\; \frac{\hbar}{T_{\mathrm{loop}}},
\la{est-1}
\end{equation}
and
\begin{equation}
\delta_{ E} \equiv |\mathcal E_{E^{(1)}}-\mathcal E_{E^{(2)}}| \;\ll\; \frac{\hbar}{T_{\mathrm{loop}}}.
\la{est-2}
\end{equation}
These conditions isolate the $E$–doublet adiabatically from higher vibrational
modes while keeping it effectively degenerate over the loop time
$T_{\mathrm{loop}}$. In this energy regime, non-Abelian transport of the
$E$–doublet along a closed loop on Kendall’s shape sphere $\mathbb S^{2}_{K}$ is
governed by the SU(2) Wilczek-Zee connection $\mathcal A$.

\subsubsection{Control knobs and mapping to $(A,\psi)$:}

The complex control parameter $\psi=\rho+i\eta$ in (\ref{psi})  is controlled by 
synchronized, low-amplitude modulations of the two
base bond lengths $r_{13} $ and $ r_{23}$ realized via acousto-optic control of the tweezer
positions and depths.  We map each point $(\theta,\phi)$ on Kendall's shape sphere $\mathbb{S}^2_K$ to a 
Bloch vector $\mathbf{n}(\theta,\phi)$ for the $E$-doublet, with the corresponding instantaneous 
eigenbasis defined up to local $\mathrm{U}(1)$ phase rotations about $\mathbf{n}$. 
We choose a gauge where $\mathbf{n}$ serves as the local quantization axis,  such that it  is 
aligned with the body-frame normal of a canonical representative of each shape.
The differentials $d\mathbf n$ and $\mathbf n\times d\mathbf n$ then 
describe how this Bloch-vector field varies over $\mathbb S^{2}_{K}$, and encode the
shape-dependent evolution of the $E$–doublet along the loop in shape space, in a manner 
consistent with zero total mechanical angular momentum.

A weak bias field defines a reference quantization axis in the laboratory frame
that we align with the equilibrium normal of the trimer. This choice
fixes the relation between the lab frame and the body frame, so that the
Berry–Wilczek–Zee evolution of the $E$–doublet is generated by the driven shape
dynamics. To suppress excitation of the breathing mode and to keep the triangle
close to fixed size so that its shape remains on $\mathbb S^2_K$,we co-modulate  the apex bond
length $r_{12}(t)$ so that
\[
\delta r_{13} (t) + \delta r_{23} (t) + \delta r_{12} (t) \simeq 0.
\]
In line with ~(\ref{nd}),  a convenient parametrization of a
small elliptical loop in shape space is
\[
\begin{matrix}
\delta r_{13}(t) & = & \epsilon  \cos(\Omega t) \\
\delta r_{23}(t) & = &\epsilon  \sin(\Omega t+\phi) \end{matrix} 
\]
with $\epsilon\ll1$, loop frequency $\Omega=2\pi/T_{\mathrm{loop}}$, and the
controllable phase $\phi$ corresponds to the relative phase
$\phi_{13}-\phi_{23}$; see also Figure~1 panel d.

We introduce the instantaneous eigenbasis of
$\mathbf n\cdot\boldsymbol\sigma$ to decompose  SU(2) connection $\mathcal A$ 
into a diagonal Abelian part $C\equiv A+\omega$ and an off-diagonal transverse
part $J\,\sigma^{+}+J^{*}\sigma^{-}$ as in (\ref{fad-1})-(\ref{fad-3}).  In the present illustrative regime the weak
bias field keeps the relevant shapes near a single patch of $\mathbb S^2_K$, so that
$\omega$ is small and $C\simeq A$, the embedded Guichardet connection. The
holonomy is then governed by the loop area on Kendall’s $\mathbb S^2_K$ through $A$,
together with the complex control parameter $\psi$ through the transverse
coupling $J$. The magnitude $|\psi|$ sets the strength, and the phase
$\arg\psi$ sets the direction, of the non-Abelian mixing within the $E$–doublet.

\subsubsection{A demonstrator loop and expected gate:}

With the normal $\mathbf n$ pinned so that $C\simeq A$, we consider a small
elliptical gate loop $\Gamma$ that encircles a solid angle
$\Omega_{\Gamma}\ll 1$ on $\mathbb S^{2}_{K}$. Using (\ref{dtheta2}), (\ref{n-ap})
the loop integral (\ref{finalA}) evaluates to
\[
\oint_{\Gamma} A = \tfrac{1}{2}\,\Omega_{\Gamma},
\]
so that the geometric gate angle is (\ref{nots})
\[
\Theta \approx \frac{q}{2}\,\Omega_{\Gamma}
+ \mathcal{O}(|\psi|^{2}),
\]
with the $\mathcal{O}(|\psi|^{2})$ corrections due to the transverse
off-diagonal components (\ref{fad-3}). Since a single small elliptical loop
produces only a small rotation angle $\Theta$, larger angles such as a
$\pi/2$ single-qubit rotation (a $\sqrt{X}$-type gate) can be realized by
repeating the loop $N$ times. This yields
\[
\Theta_{\mathrm{tot}}\approx N \frac{q}{2}\,\Omega_{\Gamma},
\]
provided the adiabaticity condition
$\delta_{E} \ll \hbar/T_{\mathrm{loop}} \ll \Delta_{\mathrm{gap}}$ remains
satisfied for each traversal.

\subsubsection{Timing and adiabaticity:}
Adiabatic transport within the $E$-manifold requires the separation of scales as
specified in (\ref{est-1}) and~(\ref{est-2}),
\begin{equation}
\delta_{E} \ll \frac{\hbar}{T_{\mathrm{loop}}} \ll \Delta_{\mathrm{gap}}.
\end{equation}
Here  $T_{\mathrm{loop}}$ is the duration of a single small loop $\Gamma$ on
Kendall’s shape sphere $\mathbb S^{2}_{K}$. The lower inequality ensures that the two
$E$-levels remain effectively degenerate over the course of one loop, while the
upper inequality suppresses nonadiabatic transitions out of the $E$-subspace.
In this regime, treating the off-diagonal couplings to the breathing mode
perturbatively yields  in the  leading order the Dyson estimate
\begin{equation}
P_{\mathrm{leak}} \sim
\left(\frac{\hbar}{T_{\mathrm{loop}}\Delta_{\mathrm{gap}}}\right)^{2},
\la{Pro}
\end{equation}
for the leakage probability $P_{\mathrm{leak}}$ out of the $E$-manifold.
When a larger rotation angle is realized by repeating small loops $N$ times, the
total gate time becomes $T_{\mathrm{gate}} = N T_{\mathrm{loop}}$. With
room-temperature Rydberg lifetimes in the few-to-few tens of
microseconds~\cite{SaffmanRMP2010}, choosing $T_{\mathrm{loop}}$ in the
sub-microsecond to few-microsecond range and $N=\mathcal{O}(1\!-\!10)$ leads to
typical total gate times in the $5$--$25~\mu\mathrm{s}$ range. This is compatible
with the Cs($nd_{3/2}$) Rydberg lifetime, especially under cryogenic
conditions. While such gate durations are adequate for a proof-of-principle
demonstrator, achieving high-fidelity quantum-logic operation would require
either longer lifetimes, faster loops, or larger single-loop areas that reduce
$N$. In practice, axial confinement and field control, which are standard in
tweezer arrays~\cite{SaffmanRMP2010,BrowaeysLahaye2020}, provide the primary
experimental knobs for tuning $\Delta_{\mathrm{gap}}$ and $\delta_{E}$ in
Cs($nd_{3/2}$) trimers.

\subsubsection{Readout: gauge-invariant Wilson-loop trace}

A minimal readout protocol is a Ramsey/echo sequence within the $E$-manifold,
that proceeds as follows:

\vskip 0.2cm
\noindent
(i) prepare an $E$-superposition;

\noindent
(ii) execute the loop $\Gamma$ during the first free-precession window;

\noindent
(iii) apply a $\pi$-pulse that exchanges
$|E^{(1)}\rangle\!\leftrightarrow\!|E^{(2)}\rangle$;

\noindent
(iv) execute the time-reversed loop $\Gamma^{-1}$;

\noindent
(v) close with a $\pi/2$ pulse.

\vskip 0.2cm
This sequence cancels dynamical phases associated with the residual splitting
$\delta_{E}$ while doubling the geometric contribution of the Wilczek--Zee
holonomy. The resulting Ramsey fringe yields a basis-calibrated estimate of the
gauge-invariant Wilson-loop trace (\ref{U2}), (\ref{finalA})-(\ref{bigthe}).
The full trace can be reconstructed by
preparing two linearly independent initial superpositions within the
$E$-manifold. The geometric origin of the signal can be verified by reversing
the loop orientation, which flips the sign of the solid angle $\Omega_{\Gamma}$,
and by varying the enclosed area on Kendall’s shape sphere $\mathbb S^{2}_{K}$ while
keeping the loop duration $T_{\mathrm{loop}}$ fixed.

\subsubsection{Error channels and operating window:}

The dominant sources of error are:

\vskip 0.2cm
\noindent
(i) spontaneous decay and dephasing of the Rydberg electron over the total
gate time $T_{\mathrm{gate}}$, producing an error
$\sim 1 - e^{-T_{\mathrm{gate}}/\tau_{R}}$ with $\tau_{R}$ the lifetime of
the Cs($nd_{3/2}$) Rydberg state;

\noindent
(ii) nonadiabatic leakage out of the $E$-subspace during each loop, with a
per-loop leakage probability that scales as in equation ~(\ref{Pro});

\noindent
(iii) the residual splitting $\delta_{E}$ between $|E^{(1)}\rangle$ and $|E^{(2)}\rangle$
causes the two $E$-levels to acquire slightly different dynamical 
phases during the loop that,  if not refocused by the Ramsey/echo protocol, would
accumulate an unwanted dynamical phase of order
$\sim \delta_{E}T_{\mathrm{gate}}/\hbar$.  

\vskip 0.2cm
\noindent
Additional errors may arise from

\vskip 0.2cm
\noindent\noindent
(iv) stray electric fields that shift and mix the $E$–doublet, which can be
mitigated using standard field-nulling techniques;

\noindent
(v) trap-intensity and waveform noise that modulate the loop area
$\Omega_{\Gamma}$ and hence affect the geometric rotation angle, which can be
mitigated using phase-locked control of the tweezer waveforms.

\vskip 0.1cm

In summary, the conditions~(\ref{est-1}) and~(\ref{est-2}) define the required operating
window. Figure~1 of the main text illustrates how relative-phase control of the
bond oscillations shapes the loop on Kendall’s $\mathbb S^{2}_{K}$ and thereby sets the
resulting geometric rotation. Taken together with the natural bond-length scale
($\sim\!0.1~\mu\mathrm{m}$), the accessible Rydberg lifetimes, and standard
tweezer and field controls~\cite{SaffmanRMP2010,BrowaeysLahaye2020}, the present
considerations place the proposed demonstrator well within reach of current
neutral-atom technology.

\end{document}